\documentclass[aps,prl,twocolumn,showpacs,superscriptaddress,groupedaddress,preprintnumbers,amsmath,amssymb]{revtex4-1}  
\usepackage{mathrsfs}
\usepackage{mathtools}
\newcommand{\be}{\begin{equation}}
\newcommand{\ee}{\end{equation}}
\newcommand{\p}{\partial}
\newcommand{\CC}{{\mathcal{C}}}
\newcommand{\CS}{{\mathcal{S}}}
\newcommand{\BR}{{\mathbb{R}}}
\newcommand{\BN}{{\mathbb{N}}}

\newcommand{\F}{{\mathscr{F}}}
\newcommand{\Tr}{{\mathrm{Tr\,}}}
\begin{document}
\title{Entangling Fractals}
\preprint{IPM/P.A-410}
\author{Amin Faraji Astaneh}
\affiliation{School of Particles and Accelerators,
Institute for Research in Fundamental Sciences (IPM),
P.O. Box 19395-5531, Tehran, Iran} 
\email{faraji@ipm.ir}

\begin{abstract} 
We use the Heat Kernel method to calculate the Entanglement Entropy for a given entangling region on a fractal. The leading divergent term of the entropy is obtained as a function of the fractal dimension as well as the walk dimension. The power of the UV cut-off parameter is (generally) a fractional number which indeed is a certain combination of these two indices. This exponent is known as the spectral dimension. We show that there is a novel $\log$ periodic oscillatory behavior in the expression of entropy which has root in the complex dimension of the fractal. We finally indicate that the Holographic calculation in a certain hyper-scaling violating bulk geometry yields the same leading term for the entanglement entropy, if one identifies the effective dimension of the hyper-scaling violating theory with the spectral dimension of the fractal. We provide more supports with comparing the behavior of the thermal entropy in terms of the temperature computed for a fractal geometry and a hyper-scaling violating theory as well.
\end{abstract} 

\maketitle

{\it{Introduction.}} A fractal is a \emph{self similar} object with a characteristic dimension known as the \emph{fractal dimension} which exceeds its topological dimension. Interestingly, these properties are closely bound up with the known characteristics of a critical system in physics. In fact, self similarity can be translated to \emph{scale invariance} at the critical point and in correspondence with the fractal dimension we would have the \emph{critical exponents} in the theories of the critical phenomena \cite{Kroger:2000wa}.

Furthermore, the fractal geometry is somehow inherent in quantum mechanics. Indeed, as a result of the uncertainty principle, the most probable trajectory of a massive quantum particle is typically a fractal of dimension two. The importance of the fractal geometry will become more apparent when we move to a relativistic quantum many body system, due to the fact that such systems admit some fractal-like behavior at the critical points. 

On the other hand, a unique feature of any quantum mechanical system is the entanglement between its degrees of freedom which can be appropriately quantified with the Entanglement Entropy.

The entanglement entropy for a subsystem is indeed the Von-Neumann entropy constructed from the reduced density operator associated to that subsystem. To be more precise, suppose that a whole system in a pure state is divided into two individual subsystems $A$ and $\bar{A}$. The reduced density operator for each subsystem will be defined as
\be
\rho_A=\Tr\!_{\bar{A}}\,\rho_{tot}\, .
\ee
Then the entanglement entropy for the subsystem $A$ reads
\be
S_{EE}(A)=-\Tr_A\, \rho_A\log \rho_A\, .
\ee

The main purpose of this study is to compute the Entanglement Entropy for a quantum many body system whose distribution of degrees of freedom makes a fractal-like pattern. Of course, the original idea of the random fluctuating entangling regions has been presented in \cite{Solodukhin:2011zr} for the first time. As motivated above we think such calculations would be interesting since quantum mechanical systems typically exhibits fractal behavior at the critical points. Technically we employ the Heat Kernel method to achieve this goal. Of course this method applies only to a free scalar field theory.

The paper is organized as follows. The first section sets out to introduce some basics of the fractals. Then we compute the entanglement entropy for a subregion on a typical fractal using the heat kernel method. We also review some aspects of the fractal's thermodynamics in this section. We highlight the $\log$ periodic oscillations in the expression of entropy and end up with a comment on the possible Holographic description and a short conclusion.

{\it{The Characterization of a Fractal.}}
A fractal is generally characterized by two independent parameters, the \emph{fractal dimension}, (${d_f}$), and the \emph{Walk dimension}, $({d_w})$.  In what follows we briefly review the physical implications of these two indices.

\emph{The Fractal (Hausdorff) dimension:}
This index actually provides a measure of the spatial scaling on the fractal. if we denote the length scale on the fractal by $\ell$, we can define ${d_f}$ as follows
\be
{d_f}=\lim_{\ell\rightarrow 0}\frac{\log V(\ell)}{\log\ell}\, ,
\ee
where by $V(\ell)$ we mean the volume of the fractal evaluated at the length scale $\ell$. Obviously, for a smooth manifold, $V(\ell)\sim \ell^d$ and thus the fractal dimension coincides with the dimensionality of spacetime.

\emph{The Walk dimension:}
Walk dimension is the index of the anomalous diffusion on the fractal in the sense that
\be
x_{r.m.s} \propto t^{1/{d_w}}.
\ee
Then since in a general case ${d_w}>2$, one may deduce that the diffusion is slower on a fractal in comparison with a smooth manifold. 

There is still another index which has very important implications although it is not independent from the formers. The so called \emph{spectral dimension}, $d_s$, is somehow telling us about the scaling properties of the eigenvalues of the differential operator in the action of the field theory. As we just pointed out, it is not an independent index and is related to the fractal dimension and the walk dimension as $d_s=2d_f/d_w$.

Henceforth, we denote a fractal with specific indices ${d_f}$, ${d_w}$ and $d_s$ by $\F^{d_f}_{d_w}$ or simply by $\F^{d_s}$. It is worth noting that in this sense $\BR^d$ is just a particular fractal, $\BR^d\sim\F^d_2$.

{\it{Entanglement Entropy on a Fractal.}}
In this section we calculate the entanglement entropy for a spatial entangling region whose boundary, $\Sigma$, is a fractal. To achieve this we employ the usual replica method.
The main challenge in this calculation is to compute the partition function or better said the effective action, $W_n$, on a manifold with conical singularity. If we could then it would be straightforward to compute the EE through the relation
\be\label{entropy definition}
S_{EE}(\Sigma)=(n\p_n-1)W_n\big\vert_{n=1}\, .
\ee
The crucial point is that the effective action can be read off from the trace of the heat kernel as follows
\be\label{effective action}
W_n=-\frac{1}{2}\int_{\epsilon^2}^\infty \frac{ds}{s}\,\Tr K_n(s)\, .
\ee
where $\epsilon$ is the regularization parameter. Now the whole problem reduces to evaluating the heat kernel on a manifold with conical singularity.
All we need to compute is
\be\label{product HK}
\Tr K_n(s)=\Tr K_{\CC_2}(s)\times\Tr K_{\F^{d_s}}(s)\, ,
\ee
where $\CC_2$ is a two dimensional cone. Parameterizing the cone with the polar coordinates $(r,\phi)$, the replication would be possible through changing the periodicity of the angular coordinate, $\phi$, to $2\pi n$. It also should be noted that in this setup $\Sigma$ is located at $r=0$.

Fortunately, there are some known estimations for the Heat Kernels on the fractals. It has been shown that for a wide range of fractals one can write the following \emph{sub-Gaussian} estimate of the heat kernel \cite{Dunne}
\be
K_{\F^{d_f}_{d_w}}(X,X';s)\approx \frac{1}{(4\pi s)^{d_f/d_w}}\exp\left[-\left(\frac{{\vert X-X'\vert}^{d_w}}{4s}\right)^{\frac{1}{{d_w}-1}}\right]\, ,
\ee
where $X$ denotes the coordinates on the fractal and $s$ is an auxiliary time coordinate. As expected, for the especial case $(d_f=d,d_w=2)$, we recover the heat kernel on $\BR^d$.

In two dimensions it is possible to compute the heat kernel on $\CC_2$, using the Sommerfeld formula, see e.g. \cite{Solodukhin:2011gn}
\be\label{Sommerfeld}
K_n(\phi,\phi';s)=K(\phi,\phi';s)+\frac{i}{4\pi n}\int_\CC d\omega \cot(\frac{\omega}{2n})K(\phi-\phi'+\omega ;s)\, ,
\ee
here we have dropped $r$ and $r'$ for simplicity. In this formula the contour $\CC$ consists of two vertical lines from $(-\pi+i\infty)$ to $(-\pi-i\infty)$ and $(\pi-i\infty)$ to $(\pi+i\infty)$. These lines intersect the real axis, between the poles of $\cot(\frac{\omega}{2n})$.
A direct calculation yields
\be
\Tr K_{\CC_2}=\frac{1}{12n}(1-n^2)\, ,
\ee
where we have dropped a volume term which corresponds to the vacuum energy.
On the other hand
\be
\Tr K_{\F^{d_s}}(A)=\frac{1}{(4\pi s)^{d_s/2}}A_s(\Sigma)\, ,
\ee
where $A_s(\Sigma)$ is the spectral area of $\Sigma$.
Substituting these into \eqref{effective action} and \eqref{entropy definition} we ultimately arrive at the following expression for the entropy. This is actually our main result in this note
\be\label{fractal entropy}
S_{EE}(\F^{d_s})=\frac{1}{6}\frac{1}{(4\pi)^{d_s/2}}\frac{A_s(\Sigma)}{d_s\epsilon^{d_s}}\, .
\ee
Expectedly, for ${d_f}=d-2$ and ${d_w}=2$ we recover the leading term of the entanglement entropy for an entangling region on $\BR^d$.

It is worth noting that applying a similar methodology one would be able to investigate the fractal's thermodynamics. To do so one needs to compute the heat kernel on the circle of the compactified Euclidean time. Then one would be able to find the thermal effective action in terms of the temperature $T=\frac{1}{\beta}=\frac{1}{2\pi R}$. What we need to evaluate is
\be\label{thermal effective action}
W_\beta=-\frac{1}{2}\int_{\epsilon^2}^\infty \frac{ds}{s}\,(\Tr K_{S^1}(s)\times \Tr K_{\F^{d_s}}(s))\, ,
\ee
Using the image method, the trace of the heat kernel on $S^1$ is found to be \cite{Hung:2014npa}
\be
\Tr K_{S^1}(s)=\frac{2\beta}{(4\pi s)^{1/2}}\sum_{n=1}^\infty e^{-\frac{\beta^2 n^2}{4s}}\, .
\ee
Putting things together we finally arrive at \cite{FTH}
\be
S_{T}(\F^{d_s})=\frac{1}{2^{d_s-1}\pi^{(3d_s+1)/2}R^{d_s}}\Gamma\left(\frac{d_s+3}{2}\right)\zeta_R(d_s+1)V_s\, ,
\ee 
where $V_s$ is the spectral volume on the fractal and $\zeta_R(d_s+1)$ denotes the Riemannian zeta function. The dependence on the temperature
\be\label{temp fractal}
S_T\sim T^{d_s}\, .
\ee
is a very interesting feature of the fractal geometries. We will get back to this when we propose our holographic recipe.

Before going ahead we would like to elaborate more on the validity of the estimation we have used in our calculations. It should be noted that the sub-gaussian form provides just lower and upper bounds for the heat kernel on the fractal. Indeed, it has been shown that the heat kernel generally admits a $\log$ periodic oscillatory behavior between these two bounds \cite{Dunne}, see also \cite{Calcagni:2011nc} . In this sense and focusing just on the desired sub-leading term of the heat kernel trace we find
\be
\Tr K_{\F^{d_s}}(s)\sim \frac{1}{(4\pi s)^{d_s/2}}\left[1+a\cos\left(\frac{b}{d_w}\,\log s+c\right)+\cdots\right]\, ,
\ee
where $a$, $b$ and $c$ are some real constants. Using this one can show that the leading term in the EE gets the following correction
\begin{widetext}
\be\label{correction}
S_{EE}(\F^{d_s})\rightarrow S_{EE}(\F^{d_s})\left(1+\frac{ad_w^2[d_s^2\cos(c+2\frac{b}{d_w}\log\epsilon)-2bd_s\sin(c+2\frac{b}{d_w}\log\epsilon)]}{4b^2+d_w^2d_s^2}\cdots\right)\, .
\ee
\end{widetext}
 
The second term in the parenthesis is a finite term which can be absorbed by a redefinition of  $\epsilon$. It is legitimate since the regulator of the auxiliary time parameter of the heat kernel is not necessarily the same as the parameter of the UV cut-off. Apart from, one may take the sub-Gaussian estimate as an averaged asymptotic form of the heat kernel on the fractal. Although such a log periodic oscillatory behavior does not affect the leading term of the entropy, it is an important unique feature of the fractals which somehow defines the fractility. In what follows carrying out an explicit example we explore this aspect of the fractal geometry in more detail.

{\it{An Explicit Example.}}
There is a well known transformation between the heat kernel trace and the zeta function which enables us to define a fractal in an alternative way. This transformation which is known as the Mellin transformation has the following form
\be
\Tr K_{\F^{d_s}}(s)=\frac{1}{2\pi i}\int_{\CC_m}dzs^{-z}\Gamma(z)\zeta_{\F^{d_s}}(z)\, ,
\ee
where $\CC_m$ is the Mellin contour which covers all the poles of the integrand \cite{Elizalde:2012zza}. As can be seen from this relation, the poles of the zeta function correspond to the small $s$ asymptotic of the heat kernel trace and thus will generate the leading terms of the entropy. A distinguishing feature of each fractal is that its geometrical zeta function admits some complex poles. In this sense on may define a fractal as a set whose geometrical zeta functions has complex poles \cite{Lapidus}. Then the spectral dimension of the fractal will be specified as \cite{Dunne}
\be\label{dimension from zeta}
d_s=2\times\text{Max}\, \Re\ (\text{poles of}\ \zeta_{\F^{d_s}}(z))\, . 
\ee

As an illustrative example, let us consider a $2+1$ dimensional field theory at $t=0$ and with a \emph{fractal string} at $x=0$ as the divided boundary. By the fractal string, $\CS$, we mean a set of intervals in real space whose lengths make a sequence. For demonstration purposes we denote a fractal string as follows
\be
\CS:\ \cup_k \{\ell_k;m_k, k\in\BN\}\, ,
\ee
where $\ell_k$ shows a decreasing sequence of the interval lengths and $m_k$ represents the multiplicity, i.e. the number of the lengths. 

As an explicit example we normalize the length of the divider boundary to one and make a \emph{Cantor string} with the following definition 
\be
\CC\CS:\ \cup_k \{3^{-k};2^{k-1}, k\in\BN\}\, .
\ee
The zeta function for such a geometry has a tower of complex poles whose real part gives \cite{Lapidus2}
\be
d_s=\log4/\log3\, ,
\ee
and the imaginary part makes a $\log$ periodic oscillatory behavior.  Therefore the leading term of the entanglement entropy reads
\be
S_{EE}(\CC\CS)=(\text{const.})\times\frac{\ell_s}{\epsilon^{\log4/\log3}}(1+a\cos(b\log\epsilon+c))+\cdots\, .
\ee
where the spectral length, $\ell_s$ tells us at which length scale one has measured the length of the Cantor string. Interestingly, for a Cantor set the spectral dimension coincides with the fractal dimension \cite{Besicovitch}, therefore one has
\be
\ell_s=1\cdot\ell^{\log4/\log3}\, .
\ee
where $\ell\approx 0$ denotes the length scale. 
Of course, this picture can alternatively be viewed as the entangling degrees of freedom which are distributed as a \emph{Cantor dust} along the entangling boundary.

{\it{A Comment on Holography.}}
There is a very intuitive prescription for the holographic calculation of the entanglement entropy. According to the \emph{Ryu-Takayanagi}'s proposal, all we need to compute is the area of a minimal surface, $\Sigma_H$, which is governed by extending $\Sigma$ into the bulk space. Then the entanglement entropy simply reads \cite{Ryu:2006bv}
\be
S_{HE}=\frac{A(\Sigma_H)}{4G_N}\, ,
\ee
where $G_N$ is the Newton constant.

A Holographic description of our problem might be possible in the context of the Holographic Hyper-scaling Violating theories. These theories are characterize by a new exponent, $\theta$, the so called hyper-scaling violating exponent. The Lorentz invariant form of the metric in $d+3$ dimensions reads 
 
\be
ds^2=r^{-\frac{-2(d-\theta+1)}{(d+1)}}(-dt^2+dr^2+\sum_{i=1}^{d-1}dx_i^2)\, ,
\ee
Roughly speaking we can say that the dual theory lives in $d_\theta=d-\theta$ effective dimensions.

Our proposal is to identify $d_\theta$ with the spectral dimension of a fractal which lives on the boundary. This seems reasonable at least when the entropy calculation is concerned. Actually, an explicit calculation of the holographic entanglement entropy yields the following leading term of the entropy \cite{Dong:2012se}-\cite{Fischler:2012ca}.

\be
S_{EE}\sim \frac{1}{d_\theta\epsilon^{d_\theta}} \, ,
\ee
which exactly coincides with the divergent part of \eqref{fractal entropy}, if we identify $d_\theta$ with $d_s$. 

Yet another support comes from the thermodynamics on the fractals. As mentioned in \eqref{temp fractal}, the temperature appears in the entropy with a scaling exponent which is actually the spectral function. This behavior is exactly the same as what we observe in the hyper-scaling violating theories. Just to match the dimensionality on both bulk gravity/thermal field theory sides we consider a hyper-scaling violating geometry in $d+2$ dimensions. Calculating the thermal entropy in such a background one finds \cite{Dong:2012se}, \cite{Alishahiha:2012qu}
\be
S_T\sim T^{d_\theta}\, .
\ee 
which again coincides with the entropy/temperature proportionality in a fractal under the assumption $d_s\sim d_\theta$.
Therefore it seems that there are sufficient witnesses to conclude that the hyper-scaling violating bulk geometry may potentially capture some information of a fractal on the boundary. This analogy leads us to conclude that the characteristic indices of the fractal can be encoded in the hyper-scaling violation exponent of the bulk geometry. 

{\it{Conclusion.}}
In this note, computing the entanglement entropy we investigate the dynamics of a quantum mechanical system with a fractal-like boundary. A good candidate for this study would be quantum percolation. We have found the entropy for a free scalar field theory using the heat kernel methods. The entanglement entropy is generally a well defined function of the fractal's exponents and naturally yields the entropy on $\BR^d$ for some special values of these indices. We believe that this result accompanied with more generalizations for more complicated cases would be very useful to extend our understanding of the behavior of quantum many body systems at the critical points and of course in characterization of the phase transitions in a critical system. This is our main objective in this study which is still in progress in some aspects. In particular, it would be interesting to calculate the universal logarithmic terms of the entanglement entropy in order to find the central charges for the fractal geometries, \cite{Calabrese:2004eu}.

Beside these, as we have mentioned, there are some evidences that there might be a holographic description for such fractal geometries. We propose that the hyper-scaling violating geometry with a certain choice of the effective dimension may stand as a good candidate for the bulk geometry. At least we have shown that with a very special tunning of the parameters in the hyper-scaling violating geometry, one can reproduce the same leading term of the entropy on the boundary fractal, as we call it. Beside that, the hyper-scaling structure of the thermal entropy/temperature proportionality seems to be the same on both sides. However, at the moment it seems just to be an analogy. The work in this direction is in progress, we hope to report  our findings later.
\begin{acknowledgements}
I would like to thank Mohsen Alishahiha, Hessamaddin Arfaei, Gianluca Calcagni, John Cardy, Amir Esmaeil Mosaffa, Shahin Rouhani, Abbas Ali Saberi, Mahmoud Safari and Sergey Solodukhin for very useful discussions and valuable comments on the draft.  This work is supported by Iranian National Science Foundation (INSF).
\end{acknowledgements}

  

\begin{thebibliography}{10}

\bibitem{Kroger:2000wa} 
  H.~Kroger,
  Phys.\ Rept.\  {\bf 323}, 81 (2000).

\bibitem{Solodukhin:2011zr} 
  S.~N.~Solodukhin,
  J.\ Phys.\ A {\bf 45}, 374024 (2012)

\bibitem{Dunne} 
{Dunne}, G.~V.,
  Journal of Physics A Mathematical General {\bf 45}, K4016 (2012)


\bibitem{Calcagni:2011nc} 
  G.~Calcagni,
  Phys.\ Rev.\ D {\bf 84}, 061501 (2011),
  JHEP {\bf 1201}, 065 (2012),
  JCAP {\bf 1312}, 041 (2013),

  M.~Arzano, G.~Calcagni, D.~Oriti and M.~Scalisi,
  Phys.\ Rev.\ D {\bf 84}, 125002 (2011)

\bibitem{Solodukhin:2011gn} 
  S.~N.~Solodukhin,
  Living Rev.\ Rel.\  {\bf 14}, 8 (2011)

\bibitem{Hung:2014npa} 
  L.~Y.~Hung, R.~C.~Myers and M.~Smolkin,
  JHEP {\bf 1410}, 178 (2014)


\bibitem{FTH} 
{{Akkermans}, E. and {Dunne}, G.~V. and {Teplyaev}, A.},
Phys.\ Rev.\ Lett.\  {\bf 105}, 230407 (2010)

\bibitem{Elizalde:2012zza} 
  E.~Elizalde,
  Lect.\ Notes Phys.\  {\bf 855}, 1 (2012).

\bibitem{Lapidus} 
{M. L. Lapidus, M. Van Frankenhuijsen},
Contemporary Mathe-
matics {\bf 237}, 87-105 (1999)

\bibitem{Lapidus2} 
{M. L. Lapidus, M. van Frankenhuysen},
Springer, New York, (2006)

\bibitem{Besicovitch} 
{A. S. Besicovitch, S. J. Taylor},
J. London Math.Soc.  {\bf 29}, 449-459 (1954)

\bibitem{Ryu:2006bv} 
  S.~Ryu and T.~Takayanagi,
  Phys.\ Rev.\ Lett.\  {\bf 96}, 181602 (2006)


\bibitem{Dong:2012se} 
  X.~Dong, S.~Harrison, S.~Kachru, G.~Torroba and H.~Wang,
  JHEP {\bf 1206}, 041 (2012)

\bibitem{Alishahiha:2012qu} 
  M.~Alishahiha, E.~O Colgain and H.~Yavartanoo,
  JHEP {\bf 1211}, 137 (2012)

\bibitem{Fischler:2012ca} 
  W.~Fischler and S.~Kundu,
  JHEP {\bf 1305}, 098 (2013)

\bibitem{Calabrese:2004eu} 
  P.~Calabrese and J.~L.~Cardy,
  J.\ Stat.\ Mech.\  {\bf 0406}, P06002 (2004)
\end{thebibliography}
\end{document}